\documentclass[twocolumn,aps,pra,superscriptaddress,showpacs,tightenlines]{revtex4-1}
\usepackage{setspace}
\usepackage{mathptmx}
\DeclareMathAlphabet{\mathcal}{OMS}{cmsy}{m}{n}
\usepackage{mathrsfs}

\usepackage[T1]{fontenc}
\usepackage{amstext}
\usepackage{amsmath}
\usepackage{graphicx}
\usepackage{esint}
\usepackage{color}
\usepackage[colorlinks, linkcolor=blue, citecolor=blue,urlcolor=blue, hyperindex,bookmarks=false,pdfstartview=FitH]{hyperref}
\usepackage{lineno}
\makeatletter
\@ifundefined{textcolor}{}
{%
 \definecolor{BLACK}{gray}{0}
\definecolor{WHITE}{gray}{1}
 \definecolor{RED}{rgb}{1,0,0}
 \definecolor{GREEN}{rgb}{0,1,0}
 \definecolor{BLUE}{rgb}{0,0,1}
 \definecolor{CYAN}{cmyk}{1,0,0,0}
 \definecolor{MAGENTA}{cmyk}{0,1,0,0}
 \definecolor{YELLOW}{cmyk}{0,0,1,0}
}
\makeatother

\begin{document}

\title{ Unified generation and fast emission of arbitrary single-photon multimode $W$ states}

\author{Juncong Zheng}
\affiliation{Hunan Key Laboratory for Micro-Nano Energy Materials and Devices and School of
Physics and Optoelectronics, Xiangtan University, Hunan 411105, China}

\author{Jie Peng}
\email[email:]{jpeng@xtu.edu.cn}
\affiliation{Hunan Key Laboratory for Micro-Nano Energy Materials and Devices and School of
Physics and Optoelectronics, Xiangtan University, Hunan 411105, China}

\author{Pinghua Tang}
\email[email:]{pinghuatang@xtu.edu.cn}
\affiliation{Hunan Key Laboratory for Micro-Nano Energy Materials and Devices and School of
Physics and Optoelectronics, Xiangtan University, Hunan 411105, China}

\author{Fei Li}
\affiliation{Department of Science Education, Hunan First Normal University, Changsha 410205, China}

\author{Na Tan}
\affiliation{Department of Fundamental courses,PLA Army Special Operations Academy,Guilin,541002,China}

\begin{abstract}
We propose a unified and deterministic scheme to generate arbitrary single-photon multimode $W$ states in circuit QED. A three-level system (qutrit) is driven by a pump-laser pulse and coupled to $N$ spatially separated resonators. The coupling strength for each spatial mode $g_i$ totally decide the generated single-photon N-mode $W$ state $\vert W_N \rangle=\frac{1}{A}\sum_{i=1}^N g_i|0_1 0_2 \cdots 1_i 0_{i+1}\cdots 0_N\rangle$, so arbitrary $\vert W_N \rangle$ can be generated just by tuning $g_i$. We could not only generate $W$ states inside resonators but also
release them into transmission lines on demand. The time and fidelity for generating (or emitting) $\vert W_N \rangle$ can both be the same for arbitrary $N$. Remarkably, $\vert W_N\rangle$ can be emitted with probability reaching $98.9\%$ in $20-50$ ns depending on parameters, comparable to the recently reported fastest two-qubit gate ($30-45$ ns). Finally, the time evolution process is convenient to control since only the pump pulse is time-dependent.
\end{abstract}

\maketitle

{\parindent 0 pt \bf INTRODUCTION} 
\vskip 2mm

{\parindent 0 pt 
Entanglement is essential to quantum information, which is widely applied in  quantum dense coding \cite{2881 (1992)}, quantum teleportation \cite{1895 (1993)}, quantum cryptography \cite{661 (1991)}, and quantum computing \cite{325 (1997)}. It was first proposed by Einstein, Podolsky, and Rosen (EPR) to challenge the completeness
of quantum mechanics \cite{777 (1935)}. There are two inequivalent categories  of tripartite entanglement state, the Greenberger-Horne-Zeilinger (GHZ) state  \cite{George Mason University press,yangcp1,yangcp2,npj} and the $W$ state \cite{062314 (2000),oz}, which could be used to prove Bell’s theorem without inequalities \cite{032108 (2002)}. Notably, the $W$ state is more robust than the GHZ state since if one particle is traced out, it retains multiparticle entanglement \cite{032108 (2002)}. 
Extending it to the multipartite case, the general form of a $W$ state is \cite{prx 2013,asaa}
 \begin{equation} \label{W}
  \vert W_{N} \rangle=\dfrac{1}{A}\sum_{i=1}^N A_i\vert 0_1 0_2\cdots 1_i 0_{i+1}\cdots 0_N \rangle
 \end{equation}
 where $ A= \sqrt{\sum A_{i}^{2}}$.  $\vert 1 \rangle$ and $\vert 0 \rangle$ represent two orthogonal states encoded in frequencies \cite{013845 (2016),070508 (2019)}, polarizations \cite{10151 (2019),077901 (2004),044302 (2002)}, or spatial modes \cite{012308 (2016),022337 (2020),jie} of photons, or qubit energy state\cite{67 (2018),042102 (2002),014302 (2006)}.}

Various schemes for preparing $W$ states have been proposed, via spontaneous
four-wave mixing \cite{013845 (2016),070508 (2019)}, polarizing beam splitter (PBS)\cite{10151 (2019),077901 (2004),044302 (2002)}, cavity QED \cite{042102 (2002),014302 (2006),054302 (2002)}, circuit QED \cite{jie,812(2015),yanxia1,yanxia2,prl 2020,circuit}, cold neutral atoms \cite{stoj}, spin system \cite{spin} and so on. However, a unified scheme to generate arbitrary $W$ states with high fidelity, speed, and feasibility is needed. First, $W$ states are useful from lower-order to higher-order, since the later 
establishes entanglement between a large number of channels, making them favourable for realization of
quantum information process. E.g., Gottesman
et al. has shown this state is beneficial in longer baseline
telescopes \cite{070503 (2012)}. However, the preparation of higher-order W-states often involves complex bulk-optical
set-ups \cite{pra 2013, science 2009, nature 2010}, and the scheme is complex. Second, the coefficient of each basis needs to be easily tunable, because sometimes we require not only the maximum entangled $W$ state where all $A_i$'s are equal \cite{keli}, but also some other types. E. g., the prototype $W$ state  $\vert  W_3 \rangle={1}/{\sqrt{3}} (\vert 100 \rangle+\vert 010 \rangle+\vert 001 \rangle) $ is not suitable for quantum teleportation \cite{267 (2003)},  while another one  $ \vert W^\prime_{3} \rangle={1}/{2} (\vert 100 \rangle+\vert 010 \rangle+\sqrt{2}\vert 001 \rangle) $ proposed by Agrawal and Pati is perfect for teleportation and superdense coding \cite{062320 (2006),10871 (2007)}. 
 It has been demonstrated that different $W$ states of identical qubits could be transformed by entanglement concentration \cite{042302 (2012),71 (2013)}, but the process is complex. Third, states must be generated with high speed and fidelity to accelerate the operation and avoid decoherence in practical quantum information processing. Besides, the scheme should be easy to realize with high experimental feasibility.

In this paper, we present a unified deterministic scheme to generate arbitrary single-photon multimode $W$ states $\vert W_N\rangle$,
satisfying the aforementioned requirements. We have a qutrit \cite{yangcp3} (e.g., a $\Lambda$ type superconducting quantum interference device (SQUID)\cite{024303(2015),064303(2006)}, transmon \cite{transmon1,transmon2,transmon3} or an Xmon \cite{xmon}) driven by pump-laser pulses and coupled to $N$ spatially separated resonators, to obtain $\vert W_N\rangle$ Eq. \eqref{W} with $A_i$ equaling the coupling strength between the $i$-th resonator mode and the qutrit, $g_i$. Hence we can generate arbitrary $W$ state just by adjusting $g_i$. We can not only create $\vert W_N\rangle$ inside resonators  through adiabatic evolution along a dark state, but also
emit it into transmission lines through dissipation. The emission probability reaches $98.9\%$ in $20-50$ ns, comparable to the fastest two-qubit gate ($30-45$ ns) recently reported \citep{ns}. The generation (or emission) time and fidelity (or probability) can both be the same for arbitrary mode numbers. Besides, we only need to tune the pump pulse during the whole time evolution process, which is easier to control than the qutrit and resonators. 

 \begin{figure}[t]
\centering
\includegraphics[scale=0.43]{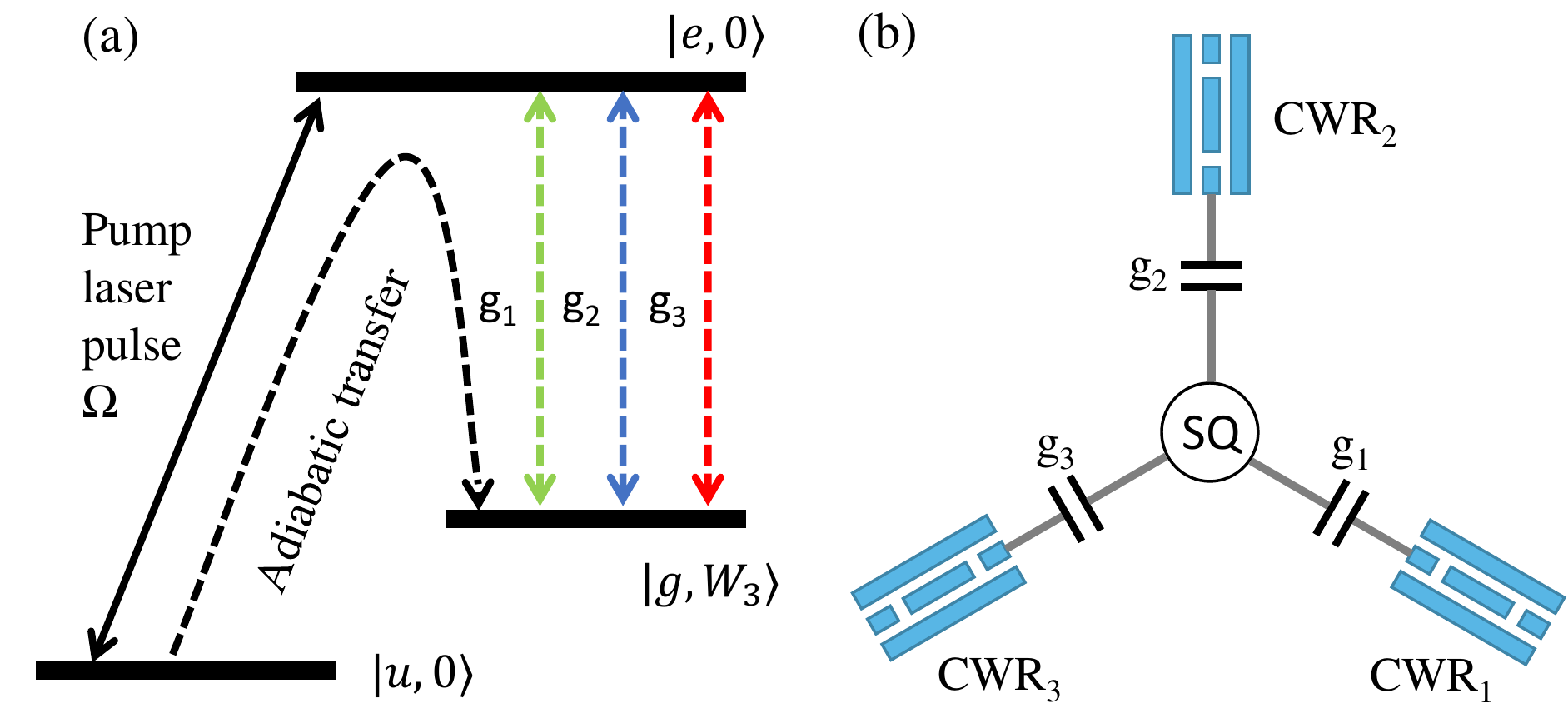}
\renewcommand\figurename{\textbf{FIG.}}
\caption[1]{\textbf{Scheme of generating the three-mode single-photon $W$ state $\vert W_3\rangle$ inside resonators through adiabatic passage.}
\textbf{a} Relevant energy levels and transitions. The
qutrit states are labeled  by $\vert u\rangle$, $\vert e\rangle$, and $\vert g\rangle$. $\vert 0\rangle$ and $\vert 1\rangle$ denote the photon
number states in the resonator. \textbf{b} Setup: A superconducting qutrit (SQ) is capacitively coupled to three coplanar waveguide resonators (CWRs) with coupling strength $g_1$, $g_2$, $g_3$ and a pump laser pulse with rabi frequency $\Omega$.}
\label{fig.1}
\end{figure}
\vskip 4mm

{\parindent 0 pt \bf RESULTS} 
\vskip 2mm

{\parindent 0 pt \bf Generating the single-photon three-mode $W$ state inside resonators}

{\parindent 0 pt 

Although we aim to generate arbitrary $N$-mode $W$ state, 
we start with the prototype three-mode case, with our our scheme depicted in Fig. \ref{fig.1}. The qutrit has a $\Lambda$-type
configuration formed by two lowest levels $|u\rangle$ and $|g\rangle$ and an excited
level $|e\rangle$. The resonator state is denoted by $\vert n\rangle$, where $n$ is the
number of photons. Three resonator modes only couple $\vert e,n\rangle$ and $\vert g,n+1\rangle$ with coupling strengths $g_1$, $g_2$, and $g_3$ respectively, whose frequencies are identical and far off resonance
from the $\vert e,n\rangle$ to $\vert u,n\rangle$ transition. Similarly, a pump laser pulse with Rabi frequency $\Omega$ only couples $\vert e,n\rangle$ to $\vert u,n\rangle$ for the same reason, as shown in Fig \ref{fig.1}. The system is initially in $\vert u,0\rangle$ and evolved into $\vert g, W_3\rangle$ through an adiabatic passage along a dark state.}

In the one-photon manifold  $\{\vert u,000\rangle,\vert e,000\rangle,\vert g,100\rangle,$\\$\vert g,010\rangle,\vert g,001\rangle\}$ under rotating wave approximation, the Hamiltonian reads $(\hbar=1)$ \cite{373 (1999),023601 (2005)}:
\begin{eqnarray}\label{eq1}
H&=&\sum_{i=u,e}E_{i}\vert i,000\rangle\langle i,000\vert +(E_g+\omega)(\vert g,100\rangle\langle g,100\vert\nonumber\\&&+\vert g,010\rangle\langle g,010\vert+\vert g,001\rangle\langle g,001\vert)\nonumber\\&&+[\dfrac{1}{2}\Omega e^{-i\omega_{P}t}\vert e,000\rangle\langle u,000\vert +g_{1}\vert e,000\rangle\langle g,100\vert\nonumber\\&&+g_{2}\vert e,000\rangle\langle g,010\vert+g_{3}\vert e,000\rangle\langle g,001\vert+H.c.] 
\end{eqnarray}
\begin{figure}[htbp]
\centering
\resizebox{1\columnwidth}{!}{
  \includegraphics{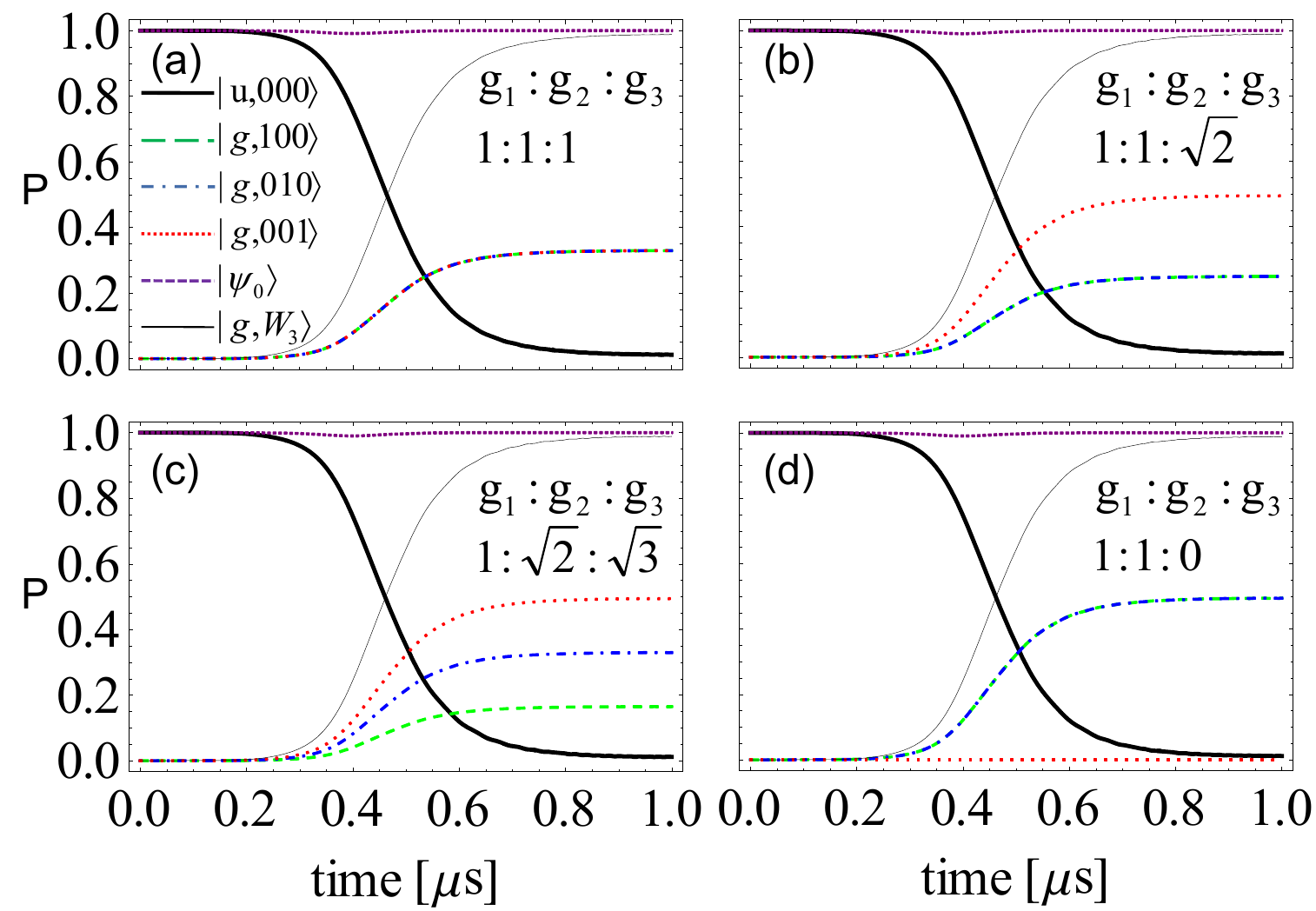}
}
\renewcommand\figurename{\textbf{FIG.}}
\caption[2]{ \textbf{Adiabatic transfer from $|u,000\rangle$ to $|g, W_3\rangle$ obtained by numerical simulation.} Population of states $\vert u,000\rangle$, $\vert g,100\rangle$, $\vert g,010\rangle$, $\vert g,001\rangle$, dark state $ \vert\Psi_0\rangle $ and final state $\vert g, W_3\rangle$, as a function of time for $\Delta_1=\Delta_2=0$ MHz, $\Omega=\Omega_0exp[-(t-\tau)^2/T_0^2]$, with $\Omega_0=2\pi\times700$ MHz, $T_0=0.36 \mu$ s, and $\tau=1\mu$ s. The couplings in  figures \textbf{a,b,c,d} are $2g_1/2\pi= 45$ MHz, $38$ MHz, $31$ MHz, $54$ MHz, respectively.}
\label{fig.2}
\end{figure}
where $ E_{i} $ represents the eigenvalue of the $i$-th energy level, $\omega $ is the frequency of each resonator. $ \omega_{P} $ is the frequency of the pump pulse. 

In the interaction picture (see 
“Methods” section), there are three degenerate dark states $\{2g_1\vert u,000\rangle-\Omega|g,100\rangle,~2g_2\vert u,000\rangle-\Omega|g,010\rangle,~2g_3\vert u,000\rangle-\Omega|g,001\rangle\}$ with constant eigenenergy $E=0$ under two-photon resonance condition, where the detunings of the pump-laser, $\Delta_1$, and the resonators, $\Delta_2$, from the respective qutrit transitions are equal \cite{373 (1999),determine}. According to the Schr\"{o}dinger equation of the system with initial state $\vert u,000\rangle$, we find the dark state in this specific case to be (see 
“Methods” section)
\begin{eqnarray}\label{dark1}
\vert\Psi_0\rangle&=&\frac{1}{{\sqrt{4A^2+\Omega^2}}}[2A\vert u,000\rangle-\Omega\vert g,W_3\rangle],
\end{eqnarray}
where $A=\sqrt{\sum_{i=1}^3 g_i^2}$, $|W_3\rangle=\frac{1}{A}(g_1\vert 100\rangle+g_2\vert 010\rangle+g_3\vert 001\rangle)$.

Since $g_i$ is adjustable, the dark state $\vert\Psi_0\rangle$ can be used to generate arbitrary single-photon three-mode $W$ states inside resonators. First we set $g_1:g_2:g_3$ on demand, and then turn on the pump pulse which rises slowly enough to ensure the adiabatic transfer from $\vert u,000\rangle$ to $\vert g\rangle\otimes\vert W_3\rangle$ when $\Omega \gg g_i$. Choosing the pump pulse $\Omega=\Omega_0exp[-(t-\tau)^2/T_0^2]$, with $\Omega_0=2\pi\times700$ MHz \cite{Pechal}, $T_0=0.36 \mu s$, and $\tau=1\mu s $, we simulate the time evolution of the system by solving the Schr\"{o}dinger equation numerically. As can be seen in Fig. \ref{fig.2}, the adiabatic fidelity \cite{145501 (2014),2433 (1995),Cambridge UK (2000)} $\vert\langle\psi(t)\vert\Psi_0\rangle\vert^2$ is close to $1$ at all time and the fidelity $F(t)=\vert\langle\psi(t)\vert g, W_3\rangle\vert^2$ increases as the pump pulse rises and finally approaches unity when $\Omega \gg g_i$. Consequently, our proposal is testified by numerical results. Ideally, $F(t)$ finally reaches $\Omega_0^2/(4A^2+\Omega_0^2)$ at $t=\tau$.  Since the highest $\Omega_0/2\pi$ we find in experiments $\approx 1$ GHz \cite{Pechal}, $A/2\pi$ can be some tens of MHz to ensure a high fidelity at the end of the adiabatic evolution. Meanwhile, the gap limiting the evolution speed is $\frac{1}{2} (\Delta_1 - \sqrt{4A^2+\Omega^2+\Delta_1^2})$. Therefore $\tau\ll A^{-1}$ \cite{373 (1999)} according to the adiabatic theorem \cite{sa}, which is limited to some hundreds of ns. Here $\tau=1\mu s$ and $F(\tau)\approx 99\%$.  Selecting specific couplings, we can generate a prototype $W$ state $1/\sqrt{3}(\vert 100\rangle+\vert 010\rangle+\vert 001\rangle)$, shown in Fig. \ref{fig.2}a, a perfect $W$ state $1/2(\vert 100\rangle+\vert 010\rangle+\sqrt{2}\vert 001\rangle)$  in Fig. \ref{fig.2}b, a common one $1/\sqrt{6}(\vert 100\rangle+1/\sqrt{2}\vert 010\rangle+\sqrt{3}\vert 001\rangle)$ in Fig. \ref{fig.2}c and a Bell state $1/\sqrt{2}(\vert 10\rangle+\vert 01\rangle)$ in Fig. \ref{fig.2}d.
      
\vskip 2mm

{\parindent 0 pt \bf Emission of the single-photon three-mode $W$ states}

{\parindent 0 pt
 \begin{figure}[htbp]
\centering
\resizebox{1\columnwidth}{!}{
  \includegraphics{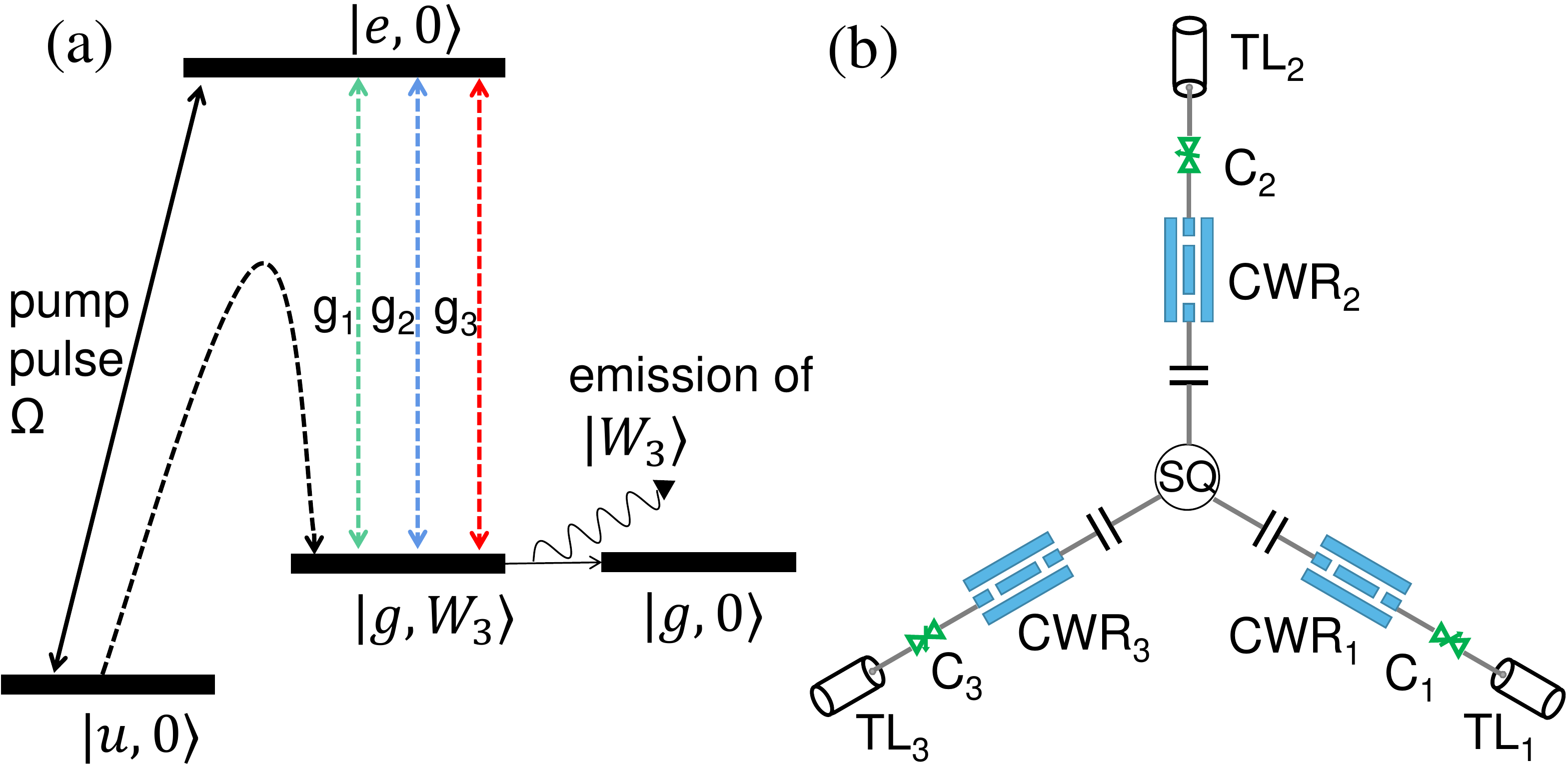}
}
\renewcommand\figurename{\textbf{FIG.}}
\caption[3]{\textbf{The scheme to emit the single-photon three-mode $W$ state $\vert W_3\rangle$.} (a) Relevant energy levels and transitions. (b) Setup: We add a variable coupler to control the dissipation rate of each CWR into the corresponding transmission line (TL) \cite{yin}.}
\label{fig.3}
\end{figure} 
Our scheme to emit the single photon $W$ state is shown in Fig. \ref{fig.3}. Each CWR is coupled to a transmission line (TL) through a variable coupler C, so that its dissipation rate $\kappa$ into the TL is tunable \cite{yin}. We are able to generate a single photon $W$ state inside CWRs when C is turned off, as discussed above. If we turn on each C with the same $\kappa$, then the $W$ state is released into the TLs, overcoming its disadvantage of uneasy to be detected \cite{054302 (2002)}.   First, we give an intuitive explanation. Any
initial superposition state of the form
\begin{equation}
g_1\vert g, 100\rangle+g_2\vert g,010\rangle+g_3\vert g,001\rangle
\end{equation}
will be finally transformed to
\begin{equation}
\vert g,000\rangle\otimes(g_1\vert 100\rangle_{out}+g_2\vert 010\rangle_{out}+g_3\vert 001\rangle_{out})
\end{equation}
through leakage resonators with the same dissipation rate, where $\vert 100\rangle_{out},\vert 010\rangle_{out},\vert 010\rangle_{out}$ denote there is one photon in the output channel of the first, second and third resonators respectively, considering 
$a_{iout}(t)=\sqrt{\kappa_i}a_{i}(t),~~~i=1,2,3$ \cite{Pechal,input} in the Heisenberg picture. A more rigorous analysis is shown in “Methods” section. }

\begin{figure}[htbp]
\centering
\resizebox{1\columnwidth}{!}{
  \includegraphics{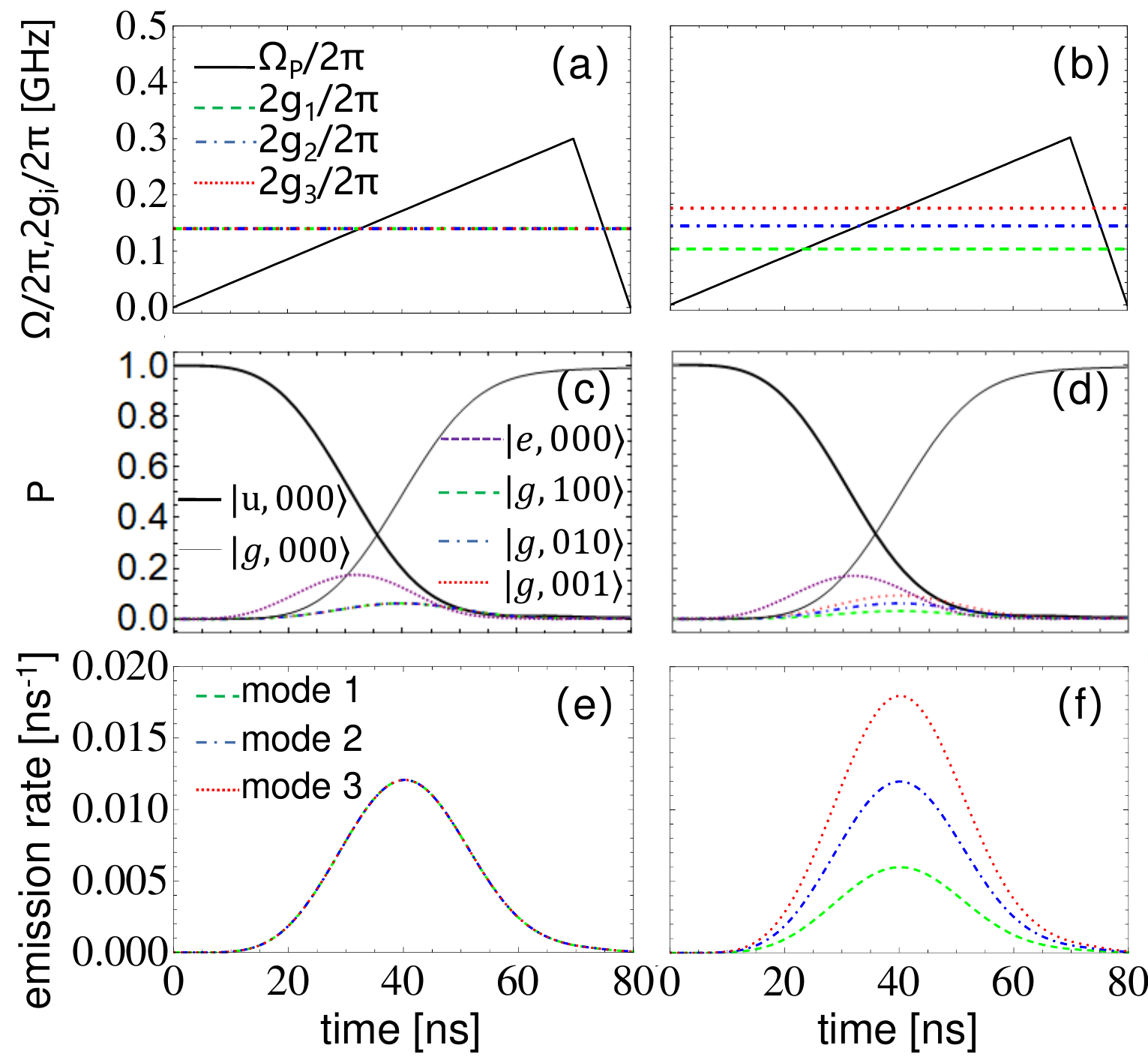}
}
\renewcommand\figurename{\textbf{FIG.}}
\caption[4]{\textbf{Simulated emission of the $W$ state using the Lindblad master equation \eqref{master}.} $\vert W_3\rangle=\frac{1}{\sqrt{3}}(\vert 100\rangle+\vert 010\rangle+\vert 001\rangle)$ in the left panel and $\frac{1}{\sqrt{6}}(\vert 100\rangle+\sqrt{2}\vert 010\rangle+\sqrt{3}\vert 001\rangle)$ in the right panel, for $\Delta_1=\Delta_2=2\pi\times 20$ MHz, $ \gamma=\gamma_{\phi}=2\pi\times 0.04$ MHz, $ \kappa_i=\kappa=0.2$ GHz. (a) and (b) Pump pulses and coupling strength between qutrit and each resonator. \textbf{c, d} Population of states $\vert u,000\rangle$, $\vert g,100\rangle$, $\vert g,010\rangle$, $\vert g,001\rangle$, $\vert g,000\rangle$ inside resonators. \textbf{e, f} Emission rate of each mode into the transmission line.}
\label{fig.4}
\end{figure}

In realistic imperfect experimental setups, $\kappa_i$ includes a intrinsic part and another one caused by the coupler $C$. The former is neglected here because it can be chosen as $1/1000$ of the later in experiment \cite{yin}. The qutrit excited state $| e\rangle$  possesses a lifetime $(2\gamma)^{-1}$, where we have assumed its decay rates to $| u\rangle$ and $| g\rangle$ are equal, and a dephasing time $(\gamma_\phi)^{-1}$.
Choosing the pump pulse $\Omega$ and coupling $g_i$ as shown in Fig. \ref{fig.4}a, b, the population transfer under dissipation are obtained by solving the Lindblad master equation (see 
“Methods” section) numerically, and shown in Fig. \ref{fig.4}c, d. The single-photon $W$ state is created at the rising edge of the pump pulse and emitted through the leakage resonators, so the system will end up with a product state $\vert g,000\rangle$. Choosing $\kappa_i=\kappa=2\pi\times0.2$ GHz \cite{yin},
$\gamma=\gamma_{\phi}=2\pi\times 0.04$ MHz \cite{blas}, $\Omega_{max}/2\pi=0.3$ GHz \cite{prl 2009}, which are within the reach of experiment, the $W$ states can be emitted in $80$ ns with probabilities reaching $99.1\%$, as shown in Fig. \ref{fig.4}e, f.  The emission rate and probability of the $i$-th resonator are proportional to $g_i^2$ for $\kappa_i=\kappa$, which means that the output single-photon state could still possesses the structure of $g_1\vert 100\rangle+g_2\vert 010\rangle+g_3 \vert 001\rangle$, an easily tunable $W$ state. This process is fast and the population of $\vert e,000\rangle$ is not always zero, as can be seen in Fig. \ref{fig.4}c, d. However, this will not hinder the emission of $\vert W_3\rangle$ since the damping rate of $\vert e\rangle$ is very small. Its population will be transferred to $\vert g,1\rangle$, and then the single photon will still be emitted through resonator dissipation.  
\begin{figure}[htbp]
\centering
\resizebox{1\columnwidth}{!}{
  \includegraphics{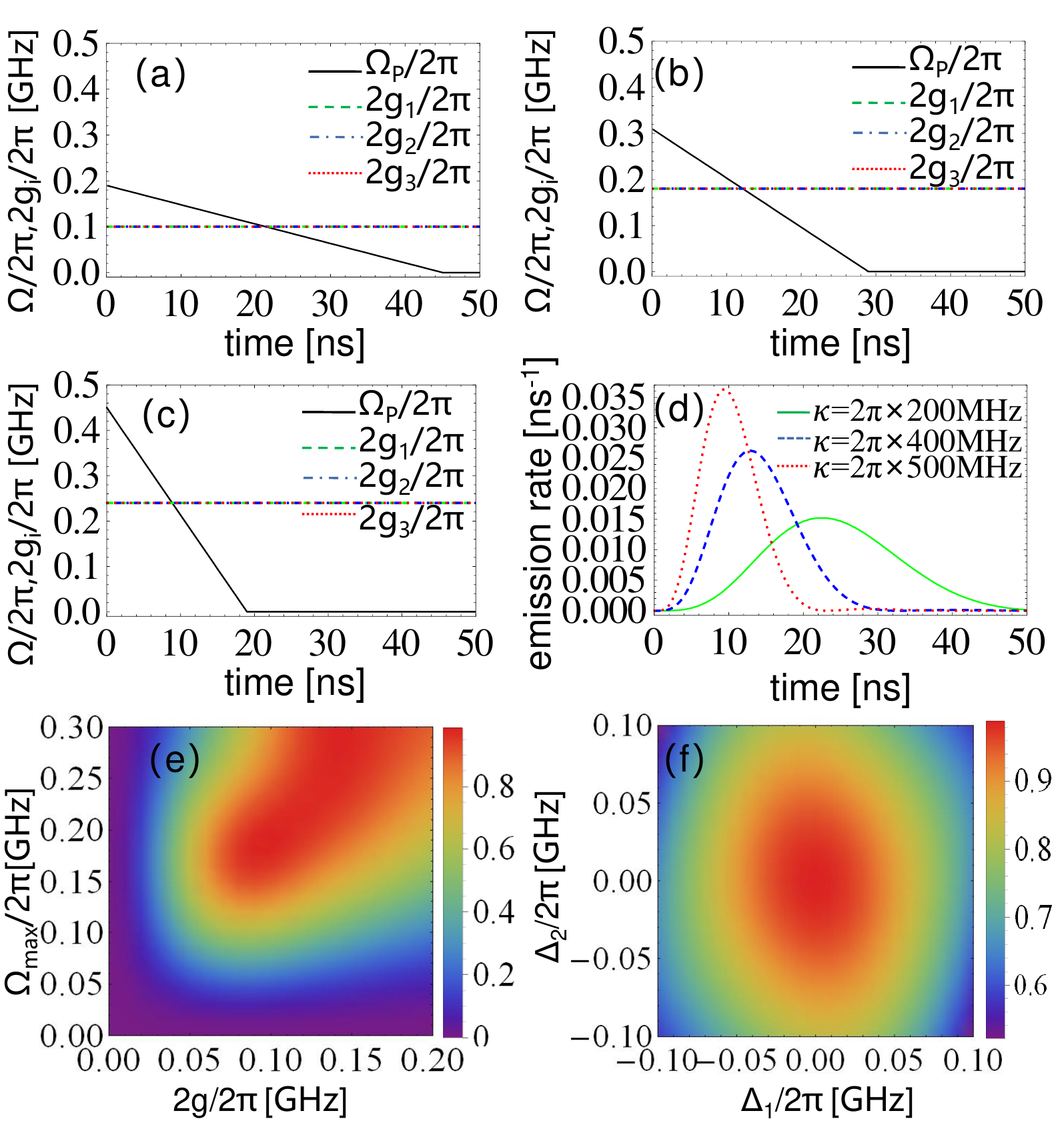}
}
\renewcommand\figurename{\textbf{FIG.}}
\caption[5]{\textbf{Simulated fast emission of the $W$ state using certain combination of parameters.} \textbf{a, b, c} $g_i$ and $\Omega$ used for $\kappa/2\pi=200$ MHz, $400$ MHz, $500$ MHz in numerical simulation, respectively. \textbf{d} The photon wave packets of each mode, which are overlapped since the emission rates are equal. \textbf{e} Emission probabilities in $50$ ns against $g$ and $\Omega_{max}$, for $\kappa/2\pi=200$ MHz,  $\gamma=\gamma_{\phi}=2\pi\times 0.04$ MHz, and  $\Delta_1=\Delta_2=0$. \textbf{f} Emission probabilities in $50$ ns against $\Delta_1$ and $\Delta_2$, for $\kappa/2\pi=200$ MHz, $\Omega$ and $g$ chosen as \textbf{a}, and  $\gamma=\gamma_{\phi}=2\pi\times 0.04$ MHz.}
\label{fig.5}
\end{figure}

The emission rate and probability depend on $g_i$, $\kappa$, $\gamma$, $\gamma_\phi$, $\Delta_1$, $\Delta_2$ and $\Omega$, and we shall explorer a combination of parameters available for fast rate and high probability. For simplicity, we consider $g=g_1=g_2=g_3$, so that the photon emission rate for three resonators are the same and their wave packets are overlapped. The maximum $\kappa$ we found in experiment is $2\pi\times200$ MHz \cite{yin}, but in principle, it could be larger in the bad-cavity limit \cite{blas,law,nc}. So we consider
$\kappa/2\pi=200$ MHz, $400$ MHz, $500$ MHz, and choose different combinations of $\Omega$ and $g$ for the fast emission of the $W$ state, which are shown in Fig. \ref{fig.5}a, b, c respectively. The corresponding  emission probabilities of the prototype $W$ state  reach $98.9\%$ in $50$ ns, $99.5\%$ in $30$ ns and $98.9\% $ in $20$ ns, respectively, as shown in Fig. \ref{fig.5}d, comparable to the recently reported fastest two-qubit gate ($30-45$ ns) \cite{ns}. Here $\gamma=\gamma_{\phi}$ is fixed at $2\pi\times 0.04$ MHz \cite{blas}, and  $\Delta_1=\Delta_2=0$. These parameters are chosen based on analyses and numerical experiments. First, as can be seen, a combination of larger $\kappa$, $g$ and $\Omega$ will make the photon wave packet sharper and shifted towards earlier time.  The system is initially in $\vert u,0\rangle$, so a large $\Omega$ will transfer the population to $|e,0\rangle$ quickly. Meanwhile, a proper combination of $g$ and $\kappa$ will quickly transfer the population of $|e,0\rangle$ to $|g,1\rangle$ and release the photon. $\Omega$ is supposed to decrease as the population of $|u\rangle$ decreases to avoid too much repumping from $|e\rangle$ to $|u\rangle$ and enhance the proportion of $|e,0\rangle\rightarrow|g,1\rangle$ transition.  Therefore, we choose a linear decreasing shape of the drive pulse, which gives the fastest emission rate in our numerical tests. We limit the total evolution time $T=50$ ns, and $\Omega(t)=\Omega_{max}(1-t/t_f)$. Taking $\kappa/2\pi=200$ MHz for example, we choose $t_f=45$ ns, and search for the best combination of $g_i$ and $\Omega_{max}$ to give the maximum emission probability numerically, as shown in Fig. \ref{fig.5}e. The best choice ($2g/2\pi=0.1$ GHz, $\Omega_{max}/2\pi=0.19$ GHz) is depicted in Fig. \ref{fig.5} a. Second, damping $\gamma$ and dephasing $\gamma_{\phi}$ will reduce the emission probability, so we choose very small $\gamma$ and $\gamma_{\phi}$ available for superconducting qubits \cite{blas}. Last, for the case shown in Fig. \ref{fig.5}a with $\kappa/2\pi=200$ MHz, we change parameters $\Delta_1$ and $\Delta_2$ to find the maximum emission probability reach at the resonance condition $\Delta_1=\Delta_2=0$, as shown in Fig. \ref{fig.5}f.

\vskip 2mm
 
{\parindent 0 pt \bf Generation and emission of arbitrary single-photon multimode $W$ states with the same fidelity and time}

{\parindent 0 pt
Here we extend our scheme to generate and emit arbitrary single-photon multimode $W$ states $\vert W_{N} \rangle=\frac{1}{A}\sum A_i\vert 0_1 0_2\cdots 1_i 0_{i+1}\cdots 0_N \rangle$. 
 The system is in principle the same as Fig. \ref{fig.3}, just with resonator number added to $N$.}

The Hamiltonian of this system becomes
\begin{eqnarray}\label{nmode}
H &=&\sum_{j=u,e,g}E_{j}\vert j,00...0\rangle\langle j,00...0\vert\nonumber\\
 &&+(E_g+\omega)(\sum_{i=1}^N \vert g, 0_1 0_2\cdots 1_i 0_{i+1}\cdots 0_N \rangle\nonumber\\&&\otimes\langle g,0_1 0_2\cdots 1_i 0_{i+1}\cdots 0_N \vert)\\&&+[\dfrac{1}{2}\Omega_{T}e^{-i\omega_{P}t}\vert e,00...0\rangle\langle u,00...0\vert \nonumber\\&&+\sum_{i=1}^N g_i\vert g, 0_1 0_2\cdots 1_i 0_{i+1}\cdots 0_N \rangle\langle e,00...0\vert+H.c.] \nonumber.
\end{eqnarray}

First we consider the generation of $W$ states inside resonators. Supposing the resonator dissipation, qutrit damping and dephasing are negligible, and the system is in state $\vert\psi\rangle=c_u|u,000\rangle+c_e|e,000\rangle+\sum_{i=1}^N c_i\vert 0_1 0_2\cdots 1_i 0_{i+1}\cdots 0_N \rangle$. Similar to the three-mode case, we solve the eigenenergy equation in the interaction picture and find $N$ degenerate dark states $\{2g_1\vert u,00\ldots0\rangle-\Omega|g,10\ldots0\rangle,~2g_2\vert u,00\ldots0\rangle-\Omega|g,01\ldots0\rangle,\ldots,~2g_N\vert u,000\rangle-\Omega|g,00\ldots1\rangle\}$ with $E=0$. On the other hand, according to the sch\"{o}dinger equation $\dot{c}_i=ig_ic_e$ with initial condition $c_u=1$,  $c_i(t)/c_j(t)= g_i/g_j$.
Hence the dark state in this specific case reads:
\begin{eqnarray}\label{nm}
\vert\Psi_0\rangle&=&\frac{1}{{\sqrt{4A^2+\Omega^2}}}[2A\vert u,000\rangle-\Omega\vert g,W_N\rangle],
\end{eqnarray}
where $|W_N\rangle=\frac{1}{A}\sum_{i=1}^N g_i\vert 0_1 0_2\cdots 1_i 0_{i+1}\cdots 0_N \rangle$, $A=\sqrt{\sum_{i=1}^Ng_i^2}$.
To create an arbitrary $W$ state $\vert W_N \rangle$, we need to set $g_i$ as required and then turn on the pump pulse which rises slowly to adiabatically transfer $ \vert u,00...0\rangle$ to the target state. We shall explore a combination of parameters which gives high fidelity and speed available for all mode numbers, as we have done for three-mode case above. 
  
\begin{figure}[htbp]
\centering
\resizebox{1\columnwidth}{!}{
  \includegraphics{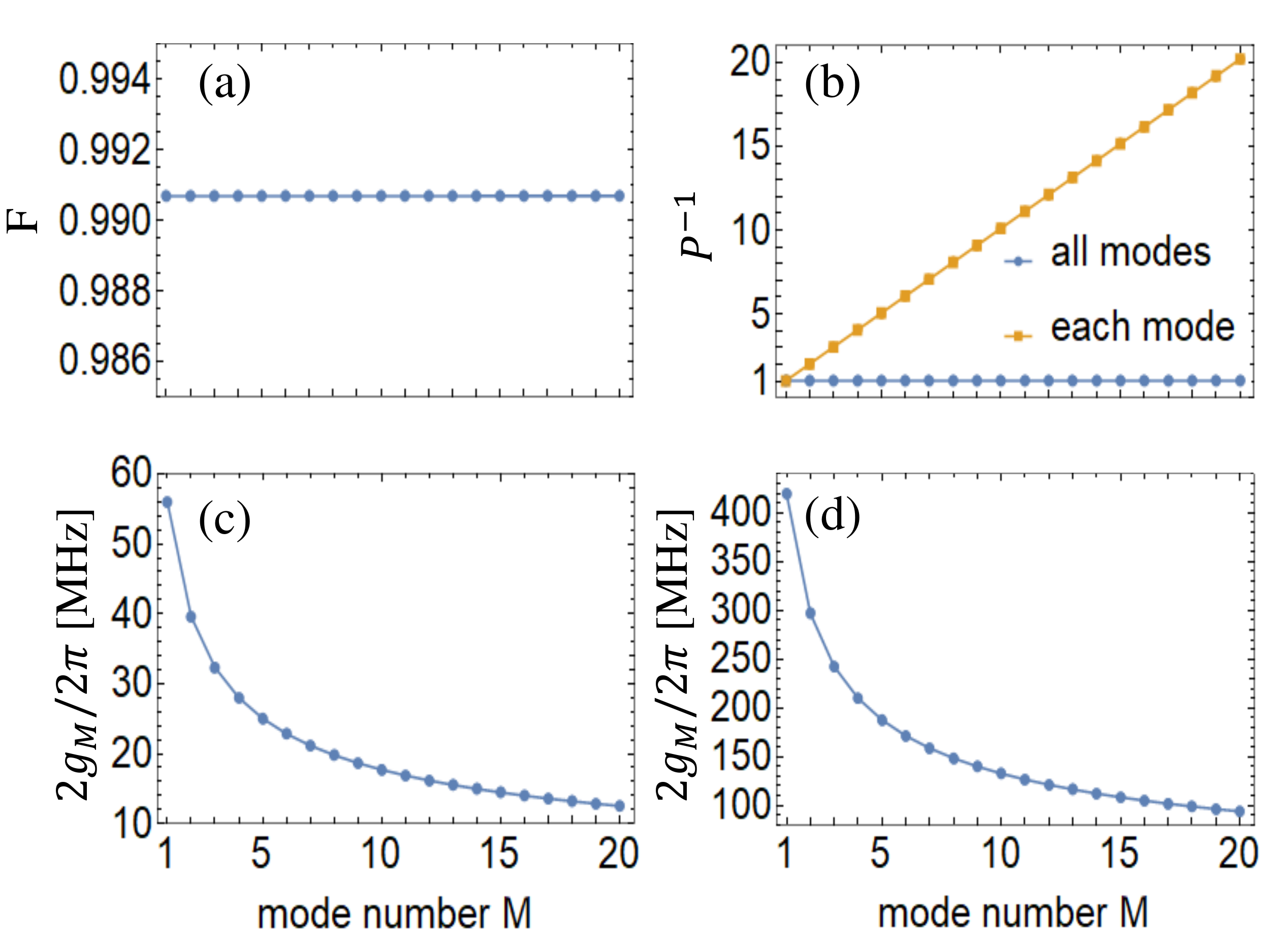}
}
\renewcommand\figurename{\textbf{FIG.}}
\caption[7]{\textbf{Fidelities of $|W_M\rangle=\frac{1}{M}\sum_{i=1}^M |0_1 0_2 \cdots 1_i 0_{i+1}\cdots 0_M\rangle$ within the same evolution time for different $M$.} \textbf{a} Fidelities of generating $|W_M\rangle$ inside resonators in $1$ $\mu$s with $M$ ranging from $1$ to $20$. $\Delta_1=\Delta_2=0$ MHz , $\Omega=\Omega_0exp[-(t-\tau)^2/T_0^2]$, with $\Omega_0/2\pi=700$ MHz, $T_0=0.36 \mu$s, and $\tau=1\mu$s. \textbf{b} Reciprocal of emission probabilities of $|W_M\rangle$ (all modes and each mode) into transmission lines in $20$ ns for different $M$.  $\kappa/2\pi=500$ MHz. $\Delta_1=\Delta_2=0$ MHz. $\Omega$ is chosen as Fig. \ref{fig.5}c, and  $\gamma=\gamma_{\phi}=2\pi\times 0.04$ MHz. \textbf{c} The couplings 
 used to obtain \textbf{a}. $g_M=g_{1}/\sqrt{M}$ with $2g_1/2\pi=56$ MHz. \textbf{d} The couplings used to obtain \textbf{b}. $g_M=g_{1}/\sqrt{M}$ with $2g_1/2\pi=420$ MHz.
}
\label{fig.6}
\end{figure}

The ideal fidelity $F=\vert\langle g,W_N\vert\Psi_0\rangle\vert^2=\Omega^2/(4A^2+\Omega^2)$, meanwhile, the energy gap limiting the adiabatic speed is $\frac{1}{2} (\Delta_1 - \sqrt{4A^2+\Omega^2+\Delta_1^2})$, so intuitively, if $A=\sqrt{\sum_{i=1}^Ng_i^2}$ and other parameters are fixed for different $N$, the fidelity and evolution time will both be the same. We will give a more rigorous proof for this in “Methods” section. So according to the parameters we have already found for the three-mode case, we choose the couplings as
\begin{equation}\label{emqm}
\sum_{i^\prime=1}^{N^\prime}g^{\prime^2}_{i^\prime}=\sum_{i=1}^{N}g^2_{i}
\end{equation} 
and fix other parameters for different mode number $N$ and $N^\prime$, to obtain the same fidelity within the same time. For simplicity, we assume all $g_i$'s ($i=1,2,\ldots,M$) equal to $g_M$ for $M$-mode case.  Choosing $2g_{1}/2\pi= 56$ MHz and $g_{M}=g_{1}/\sqrt{M}$ with $M$ ranging from $1$ to $20$, as depicted in Fig. \ref{fig.6}c, we find the fidelities are all equal to $99.07\%$ for each $M$ in $1 \mu$ s by solving the Sch\"{o}dinger equation numerically, shown in Fig. \ref{fig.6}a.

Next, we consider the emission of such $W$ states through dissipation. By studying the Lindblad master equation including resonator dissipation, qutrit damping and dephasing, we are able to prove the emission rate of the $i$-th resonator $\kappa\langle a_i^\dagger a_i\rangle$ is proportional to $g_i^2$ when $\kappa_i=\kappa$, and an adjustable single-photon multimode $W$ state $\vert W_N\rangle$ could be emitted (see “Methods” section). For its practical usage in quantum information processing, we need to release the $W$ states with high rate and probability. As discussed above, a three-mode $W$ state can be emitted in $20$ ns with probability reaching $98.9\% $. Now we have to find such combinations of parameters for the $N$-mode case.

By analyzing the master equation, we find the total emission rates $\kappa\sum_i\langle a_i^\dagger a_i\rangle$ are the same for different mode numbers, under the condition Eq. \eqref{emqm} with other parameters fixed (see “Methods”).
For simplicity, we choose $g_1=g_2=\ldots=g_N$, $g_{1^\prime}=g_{2^\prime}=\ldots=g_{N^\prime}$, and reduce Eq. \eqref{emqm} to $Ng_N^2
=N^\prime g_{N^\prime}^2$. Then we choose such couplings varying with mode numbers, as shown in Fig. \ref{fig.6}d, where $2g_1/2\pi=420$ MHz, and fix other parameters as in the three-mode case where $\kappa/2\pi=500$ MHz, $\gamma=\gamma_{\phi}=2\pi\times 0.04$ MHz, $\Delta_1=\Delta_2=0$ and $\Omega$ shown in Fig. \ref{fig.5}c. The total emission rates and emission probabilities are found to be equal for any mode number $M$ by numerical simulation, where the emission probability $P$ for each mode is naturally $1/M$ of the total emission probability, as shown in Fig. \ref{fig.6}b. Now we have found proper parameters to emit the $M$-mode $W$ states in $20$ ns with probability reaching $98.9\%$ for $M$ ranging from $1$ to $20$. This result can be extended to arbitrary $M$ by choosing $g_M=g_1/\sqrt{M}$ and other parameters fixed.

\vskip 4mm

{\parindent 0 pt \bf DISCUSSION} 
\vskip 2mm

{\parindent 0 pt
We proposed a unified deterministic scheme of generating and releasing arbitrary single-photon multimode $W$ state on demand with high emission rate and experimental feasibility. We have a qutrit coupled to $N$ spatially separated resonators and a pump pulse. Making the system evolve adiabatically along a dark state, we finally obtain arbitrary $W$ states inside resonators. The first merit of our scheme is the coefficient for the $i$-th basis $\vert 0_1 0_2\cdots 1_i 0_{i+1}\cdots 0_N \rangle$ of this $W$ state is proportional to the coupling strength of the $i$-th mode $g_i$, so we can obtain arbitrary $W$ state by changing this coupling. Second, we can release such $W$ states into transmission lines on demand, by adding a variable coupler to modulate the dissipation rate of the resonators. The emission probability reach $98.9\%$ in $20-50$ ns, depending on parameters, comparable to the fastest two-qubit gate. And we only need to vary the pump-laser pulse during the time evolution process, which is easier to control than the qutrit.
Third, the generation (or emission) time and fidelity (or probability) can both be the same by choosing $\sum_{i^\prime=1}^{N^\prime}g^{\prime^2}_{i^\prime}=\sum_{i=1}^{N}g^2_{i}$ and other parameters fixed for $N$-mode and $N^\prime$-mode cases. It is interesting to consider its experimental realization in circuit QED or related systems.}

\vskip 4mm

{\parindent 0 pt \bf METHODS} 
\vskip 2mm
{\parindent 0 pt \bf Dark states for generating $\vert W_3\rangle$}

{\parindent 0 pt The dark state $\vert\Psi_0\rangle$ Eq. \eqref{dark1} is obtained in 
the interaction picture with respect to free Hamiltonian $H_0=E_{u}\vert u,000\rangle\langle u,000\vert +(E_{u}+\omega_P)\vert e,000\rangle\langle e,000\vert +(E_{u}+\omega_P)(\vert g,100\rangle\langle g,100\vert+\vert g,010\rangle\langle g,010\vert+\vert g,001\rangle\langle g,001\vert)$, where the Hamiltonian Eq. \eqref{eq1} reduces to
\begin{equation}
\dfrac{1}{2} \begin{pmatrix}
0&\Omega&0&0&0\\
\Omega^*&2\Delta_1&2g_1&2g_2&2g_3\\
0&2g_1^*&2(\Delta_1-\Delta_2)&0&0\\
0&2g_2^*&0&2(\Delta_1-\Delta_2)&0\\
0&2g_3^*&0&0&2(\Delta_1-\Delta_2)\\
\end{pmatrix},
\end{equation}
where $\Delta_1=(E_{e}-E_{u})-\omega_P$ and $\Delta_2=(E_{e}-E_{g})-\omega$. Under the two-photon resonance condition $\Delta_1=\Delta_2$, there are three degenerate dark states $\{2g_1\vert u,000\rangle-\Omega|g,100\rangle,~2g_2\vert u,000\rangle-\Omega|g,010\rangle,~2g_3\vert u,000\rangle-\Omega|g,001\rangle\}$ with constant eigenenergy $E=0$, so a general solution with $E=0$ reads:
\begin{eqnarray}
\label{dark} 
\vert\Psi_0\rangle=\frac{1}{{\cal N}}[2\sum_{i=1}^3 A_i g_i\vert u,000\rangle-\Omega A\vert g, W_3\rangle],
\end{eqnarray}
where $A=\sqrt{\sum_{i=1}^3 A_i^2}$, $\vert g, W_3\rangle=\vert g\rangle\otimes\vert W_3\rangle$, $|W_3\rangle=\frac{1}{A}(A_1\vert 100\rangle+A_2\vert 010\rangle+A_3\vert 001\rangle)$ and ${\cal
N}=\sqrt{4(\sum_{i=1}^3{A_ig_i})^2+\Omega^2A^2}$ is the normalizing constant. The proposed initial state is $\vert u,000\rangle$, hence $\Omega A_i(t=0)=0$. Combined with Schr\"{o}dinger equation $\frac{{\rm d}(\Omega A_i)}{{\rm d}t}=-i g_ic_e$, where $c_e=\langle e,000|\psi\rangle$, we obtain $A_1:A_2:A_3=g_1:g_2:g_3$ in this specific situation, and reduce Eq. \eqref{dark} to $\vert\Psi_0\rangle$ Eq. \eqref{dark1}.

\vskip 2mm
{\parindent 0 pt \bf Lindblad master equation and emission of $\vert W_3\rangle$}

{\parindent 0 pt $\vert W_3\rangle$ is released through resonator dissipation, and
the dynamics of the system is governed by the Lindblad master equation \cite{epjd2010,814 (2010),369  (2002),3140  (2018), Oxford Univ. Press (2006),1379 (2015)}:
\begin{eqnarray}\label{master}
\dfrac{d\rho}{dt}=-i[H,\rho]+\sum_{i=1}^{N+2}\dfrac{1}{2}(2L_i\rho L_{i}^\dagger-L_{i}^\dagger L_i\rho-\rho L_{i}^\dagger L_i)
\end{eqnarray}
where $ L_1=\sqrt{\kappa_1}\vert g,000\rangle\langle g,100\vert $, $ L_2=\sqrt{\kappa_2}\vert g,000\rangle\langle g,010\vert $, $ L_3=\sqrt{\kappa_3}\vert g,000\rangle\langle g,001\vert $, $ L_{N+1}=\sqrt{\gamma}(\vert u,000\rangle\langle e,000\vert $+$\vert g,000\rangle\langle e,000\vert) $, $ L_{N+2}=\sqrt{\gamma_\phi}(\vert e,000\rangle\langle e,000\vert $, and $N$ is the resonator mode number which equals three here. Supposing the system is in state $\rho=\sum p_m|\psi_m\rangle\langle\psi_m\vert$ with $\vert\psi_m\rangle=c_{um}|u,000\rangle+c_{em}|e,000\rangle+c_{1m}|g,100\rangle+c_{2m}|g010\rangle+c_{3m}|g001\rangle+c_{gm}|g000\rangle$ with $c_{um}=1$ initially and $ \kappa_i=\kappa $, we find $c_{1m}:c_{2m}:c_{3m}=g_1:g_2:g_3$ is a solution to
\begin{eqnarray}
\dfrac{d (c_{1m}c^*_{1m})}{dt}&=&-ig_1(c_{em}c_{1m}^*-c_{em}^*c_{1m})-\kappa c_{1m}c^*_{1m}, \label{eq:24}\\
\dfrac{d (c_{2m}c^*_{2m})}{dt}&=&-ig_2(c_{em}c_{2m}^*-c_{em}^*c_{2m})-\kappa c_{2m}c^*_{2m},\label{eq:25}\\
\dfrac{d (c_{3m}c^*_{3m})}{dt}&=&-ig_3(c_{em}c_{3m}^*-c_{em}^*c_{3m})-\kappa c_{3m}c^*_{3m},\label{eq:26}
\end{eqnarray}
such that the master equation Eq. \eqref{master} can be satisfied.
Therefore, the emission rate of each mode $\kappa Tr( \rho a_i^\dag a_i)$ and its probability $\kappa\int Tr( \rho a_i^\dag a_i) dt$ will be proportional to $g_i^2$, and a desired $\vert W_3\rangle=\frac{1}{A}(g_1|100\rangle_{out}+g_2|010\rangle_{out}+g_3|001\rangle_{out})$ could be created in the output channels considering $a_{iout}(t)=\sqrt{\kappa}a_{i}(t)$.}

\vskip 2mm
{\parindent 0 pt \bf Condition for generating $\vert W_N\rangle$ with the same fidelity within the same time for arbitrary $N$}

According to the Hamiltonian Eq. \eqref{nmode}, the Schr\"{o}dinger equation gives
\begin{eqnarray}
\dfrac{d c_u}{dt}&=&-\frac{i}{2}\Omega c_e,\nonumber\\
\dfrac{d c_e}{dt}&=&-\frac{i}{2}(\Omega c_u+2\Delta_1 c_e+\sum_{i=1}^N 2g_i c_i),\nonumber\\
\dfrac{d c_i}{dt}&=&-ig_ic_e\nonumber.
\end{eqnarray}
It is easy to find if $\sum_{i^\prime=1}^{N^\prime}g^{\prime^2}_{i^\prime}=\sum_{i=1}^{N}g^2_{i}$, while other parameters are fixed
for mode numbers $N$ and $N^\prime$, then 
\begin{eqnarray}\label{ij1}
c^\prime_e(t)&=&c_e(t),\\
c^\prime_u(t)&=&c_u(t),\\\label{ij2}
c_i(t)/c_{j}(t)&=&g_i/g_{j}\\\label{ij3}
c_{i^\prime}^\prime(t)/c_{i}(t)&=&g_{i^\prime}^\prime/g_{i}\label{ij4}
\end{eqnarray} 
are solutions of the Schr\"{o}dinger equation for $N$ and $N^\prime$, where $c^\prime_e(t)$, $c^\prime_u(t)$ and $c^\prime_{i^\prime}(t)$ are corresponding wave functions for $N^\prime$ case.  Therefore, the fidelity to obtain $|W_N\rangle$ and  $|W_{N^\prime}\rangle$ are the same at any time and once we find a combination of parameters to generate a $N$-mode W state with high speed and fidelity, we can easily obtain such parameters for arbitrary $N^\prime$ by using Eq. \eqref{emqm}.}

Then we consider the emission of $\vert W_N\rangle$ using the master equation Eq. \eqref{master} with $N$ extended to arbitrary positive integers. First we assume the system is in state  $\rho=\sum p_m|\psi_m\rangle\langle\psi_m\vert$ with $\vert\psi_m\rangle=c_{um}|u,00\ldots0\rangle+c_{em}|e,00\ldots0\rangle+\sum_{i=1}^Nc_{im}|g,00\ldots1_i 0_{i+1}0_N\rangle+c_{gm}|g00\ldots0\rangle$ with $c_{um}=1$ initially and $\kappa_i=\kappa$. Then following the same routine as the three-mode case, we can prove the emission rate of each resonator $\kappa\langle a_i^\dagger a_i\rangle$ is proportional to $g_i^2$, and an adjustable single-photon multimode W state could be emitted. 

For each $\vert \psi_m\rangle\langle \psi_m\vert$, the master equation Eq. \eqref{master} with subscript ``m'' neglected reads 
\begin{eqnarray}
\dfrac{d (c_{i}c^*_{i})}{dt}&=&ig_i(c_e^*c_i-c_ec_i^*)-\kappa c_ic^*_i, \\
\dfrac{d (c_ec^*_e)}{dt}&=&-\frac{i}{2}[\Omega(c_{u}c^*_e-c_{u}^*c_e)+\sum_{i=1}^N 2g_i(c_ic_e^*-c_i^*c_e)]\nonumber\\
&&-2\gamma c_ec^*_e,\label{eq:27}\\
\dfrac{d (c_uc^*_u)}{dt}&=&-\frac{i}{2}\Omega(c_{e}c^*_u-c_{e}^*c_u)+\gamma c_ec^*_e,\\
\dfrac{d (c_uc^*_e)}{dt}&=&-\frac{i}{2}[\Omega(c_{e}c^*_e-c_{u}^*c_u)-\sum_{i=1}^N 2g_i c_u c_i^*-2\Delta_1c_u c^*_e]\nonumber\\&&
-(\gamma+\frac{\gamma_{\phi}}{2})c_uc^*_e,\\
\dfrac{d (c_uc^*_i)}{dt}&=&-\frac{i}{2}[\Omega c_{e}c^*_i-2g_i c_{u}c_e^*]-\frac{\kappa}{2}c_uc^*_i,\\
\dfrac{d (c_ec^*_i)}{dt}&=&-\frac{i}{2}[\Omega c_{u}c^*_i+\sum_{j=1}^N 2g_j c_j c_i^* +2\Delta_1c_e c^*_i-2g_i c_e c^*_e]\nonumber\\&&
-(\gamma+\frac{\kappa+\gamma_{\phi}}{2})c_ec^*_i,\\
\dfrac{d (c_ic^*_j)}{dt}&=&i[g_j c_{i}c^*_e-g_i c_{e}c_j^*]-\kappa c_ic^*_j.
\end{eqnarray}
We find Eqs. \eqref{ij1}--\eqref{ij4}
are still solutions of the above equation set for different mode numbers $N$ and $N^\prime$ under condition $\sum_{i^\prime=1}^{N^\prime}g^{\prime^2}_{i^\prime}=\sum_{i=1}^{N}g^2_{i}$ and other parameters fixed. The total emission rates $\kappa\sum_i^N |c_i|^2$ and $\kappa\sum_{i^\prime=1}^{N^\prime} |c^\prime_{i^\prime}|^2$ are equal according to Eqs. \eqref{emqm} and \eqref{ij4}. So in this specific case, the total emission rate and probability of $\vert W_N\rangle$ and $\vert W_{N^\prime}\rangle$ will be the same at anytime.

\vskip 4mm

{\parindent 0 pt \bf DATA AVAILABILITY}
\vskip 2mm

{\parindent 0 pt The data that support the findings of this study are available from the authors upon reasonable request.}
\vskip 4mm

{\parindent 0 pt \bf CODE AVAILABILITY}
\vskip 2mm

{\parindent 0 pt The codes that are used to produce the data presented in this study are available from the authors upon reasonable request.}

\end{document}